\documentclass[12pt]{article}
\pdfoutput=1
\usepackage{sectsty}
\usepackage{mathrsfs}
\allsectionsfont{\sffamily}

\usepackage[nosort]{cite}

\usepackage[pdftex]{graphicx}
\usepackage[font={scriptsize,singlespacing}]{caption}
\usepackage{epstopdf}
\usepackage{microtype}
\usepackage{psfrag}
\usepackage{wrapfig}

\usepackage{mathrsfs}  
\usepackage{calrsfs}

\usepackage{amsmath}
\usepackage{amssymb}
\usepackage{subfigure}
\usepackage{hyperref}
\usepackage{url}
\usepackage{xcolor}
\usepackage{color}

\definecolor{amaranth}{rgb}{0.9, 0.17, 0.31}
\definecolor{purple(munsell)}{rgb}{0.62, 0.0, 0.77}
\definecolor{americanrose}{rgb}{1.0, 0.01, 0.24}
\definecolor{palatinateblue}{rgb}{0.15, 0.23, 0.89}
\definecolor{royalblue(web)}{rgb}{0.25, 0.41, 0.88}
\definecolor{hanpurple}{rgb}{0.32, 0.09, 0.98}
\definecolor{beaublue}{rgb}{0.74, 0.83, 0.9}
\definecolor{carminered}{rgb}{1.0, 0.0, 0.22}
\definecolor{brightpink}{rgb}{1.0, 0.0, 0.5}
\definecolor{vividviolet}{rgb}{0.62, 0.0, 1.0}
\hypersetup{ linktoc=all,
    colorlinks, linkcolor={palatinateblue},
    citecolor={brightpink}, urlcolor={amaranth}}

\newcommand{\changeurlcolor}[1]{\hypersetup{urlcolor=#1}}  
  
\def\sideremark#1{\ifvmode\leavevmode\fi\vadjust{\vbox to0pt{\vss
 \hbox to 0pt{\hskip\hsize\hskip1em
 \vbox{\hsize2cm\tiny\raggedright\pretolerance10000
 \noindent #1\hfill}\hss}\vbox to8pt{\vfil}\vss}}}%
\newcommand{\bo}{\raise-1mm\hbox{\Large$\Box$}}

\newcommand{\be}{\begin{equation}}
\newcommand{\ee}{\end{equation}}
\newcommand{\bea}{\begin{eqnarray}}
\newcommand{\eea}{\end{eqnarray}}

\renewcommand{\d}[1]{\ensuremath{\operatorname{d}\!{#1}}}

\begin{document}
\thispagestyle{empty}
\begin{center}

\null \vskip-1truecm \vskip2truecm

{\LARGE{\bf \textsf{Lux in obscuro II: Photon Orbits of Extremal AdS Black Holes Revisited \vskip0.2truecm 
}}}

\vskip1truecm
\textbf{\textsf{Zi-Yu Tang}}\\
{\footnotesize\textsf{(1) School of Physics and Astronomy,\\
Shanghai Jiao Tong University, Shanghai 200240, China}\\
{\footnotesize\textsf{(2) Collaborative Innovation Center of IFSA (CICIFSA), \\Shanghai Jiao Tong University, Shanghai 200240, China
}}\\
{\tt Email: tangziyu@sjtu.edu.cn}}\\

\vskip0.4truecm
\textbf{\textsf{Yen Chin Ong \& Bin Wang}}\\
{\footnotesize\textsf{(1) School of Physics and Astronomy,\\
Shanghai Jiao Tong University, Shanghai 200240, China}\\ 
{\footnotesize\textsf{(2) Center for Gravitation and Cosmology, College of Physical Science
and Technology,\\ Yangzhou University, Yangzhou 225009, China}}\\
\vskip0.05truecm
{\tt Email: ongyenchin@gmail.com, wang\_b@sjtu.edu.cn}}\\

\end{center}
\vskip1truecm \centerline{\textsf{ABSTRACT}} \baselineskip=15pt

\medskip
A large class of spherically symmetric static extremal black hole spacetimes possesses a stable null photon sphere on their horizons.
For the extremal Kerr-Newman family, the photon sphere only really coincides with the horizon in the sense clarified by Doran.
The condition under which photon orbit is stable on an asymptotically flat extremal Kerr-Newman black hole horizon has recently been clarified;
it is found that a sufficiently large angular momentum destabilizes the photon orbit, whereas electrical charge tends to stabilize it.
We investigated the effect of a negative cosmological constant on this observation, and found the same behavior in the case of an
extremal asymptotically Kerr-Newman-AdS  black holes in $(3+1)$-dimensions. In $(2+1)$-dimensions, in the presence of electrical charge, the angular momentum never becomes large enough to destabilize the photon orbit. We comment on the instabilities of black hole spacetimes with a stable photon orbit.

\vskip0.4truecm
\hrule

\section{Introduction: Photon Orbits and Their Stability}

The curvature of spacetime around a black hole allows a massless particle to move in a closed orbit. Such a geodesic is aptly called  ``photon orbit''. For a static, spherically symmetric black hole, there is not just a single orbit, but an entire photon sphere. Photon orbits are in general unstable -- a slight perturbation can either cause the particle to plunge into the black hole, or cause it to escape to future null infinity. 

Mathematically, there are quite a few uniqueness results that either concern or involve photon orbits in one way or another \cite{u1,u2,u3,u4,u5}. In addition, photon orbits often saturate many inequalities, see, e.g. \cite{1701.06587} and \cite{1201.0068}. 
A recent example is that, in asymptotically flat $(3+1)$-dimensional spacetimes, any compact 2-surface $\Sigma$ whose mean curvature and its derivative for outward direction are positive on a spacelike hypersurface with non-negative Ricci scalar must satisfy the inequality $\text{Area}(\Sigma) \leqslant 4\pi (3M)^2$, where $M$ denotes the ADM mass of the black hole. The equality is saturated precisely when $\Sigma$ is the photon sphere in a hypersurface isometric to $t=\text{const}.$ slice of the Schwarzschild spacetime \cite{1701.00564,1704.04637}. 

Despite their generic instability, photon orbits do play significant roles in black hole physics. First of all, photon orbits are related to the geometric optics cross section. For an asymptotically flat Schwarzschild black hole with Schwarzschild radius $r_{h}=2M$, its geometric optics cross section is $27\pi M^2$, which corresponds to a sphere of radius $3\sqrt{3}M$, the impact parameter for a massless particle coming in from infinity to end up on the photon sphere. In astrophysical observations of black holes, this means that there will be a dark region, known as the ``black hole shadow" \cite{1705.07061}, which provides tentative evidence for the existence of black holes \cite{Takahashi:2004xh}. The forthcoming direct observation by the Event Horizon Telescope will capture exactly such a shadow {\cite{EHT1, EHT2, EHT3}}. An observation of the photon orbit itself would also suggest the existence of a black hole \cite{1406.5510}.
Recently, it has been emphasized that the ringdown signature in the gravitational wave during the mergers of two black holes, which has already been directly detected by LIGO \cite{Abbott:2016blz}, is associated with the existence of photon orbits \cite{Cardoso:2016rao} (see, however, \cite{1609.00083}). More specifically, the frequency and damping time of the ringdown waveforms are respectively related to the orbital frequency and the instability timescale of the photon orbits. 

Moreover, the instability of a photon orbit is related to the \emph{stability}
of a given spacetime, since an arbitrary number of photons (and other massless particles) can pile up on any stable photon orbit, which would eventually back-react on the spacetime geometry \cite{Khoo:2016xqv}. Indeed, the presence of stable photon orbits also slows down the decay rate of fields around a black hole \cite{1404.7036}  (see also the discussion in \cite{1605.07193}). Furthermore, as commented in \cite{Cvetic:2016bxi}, a less known implication of the presence of photon orbits is that they signal the possibility of a York-Hawking-Page type phase transition \cite{Y, HP}. The underlying reason is that the Dirichlet boundary-value problem in Euclidean quantum gravity might have multiple solutions \cite{0301026}, which jump in number when the boundary passes through a photon orbit. In the Discussion section, we shall see that this has possible tie-in with other results on the phase transitions and instabilities of extremal black holes. 

Surprisingly, it has been shown that a large class of spherically symmetric static extremal black hole spacetimes possesses a stable photon sphere (also known as ``anti-photon sphere" in the literature\cite{Cvetic:2016bxi}) on their horizons. These photon spheres are defined via the effective potential for massless particles (see \cite{Khoo:2016xqv} and the references therein). From the work of Pradhan and Majumdar \cite{1001.0359}, and the recent investigation by Khoo and Ong \cite{Khoo:2016xqv}, we now know that
a photon sphere can occur on the black hole horizon in the exactly extremal case. (Though, of course in general, stable photon orbits are not restricted to be on the extremal black hole horizon \cite{1605.07193}.) Assuming analyticity of the metric, Khoo and Ong proved a general theorem: for a large class of spherically symmetric static extremal black hole spacetimes, the photon orbit on its horizon is stable if the first nonzero Taylor coefficient of its metric coefficient $f^{(k)}(r_{h})/k!$, $k \geqslant 2$, is with an even $k$. 

The situation for black holes that are not spherically symmetric is far less clear. 
Once rotation sets in, the spherical symmetry is reduced to an axial symmetry. The behavior of photon trajectories on a photon sphere then becomes vastly more complicated
than the static case \cite{teo, 1210.2486}. For a fair and clear comparison with the static case, we will restrict our discussion to the equatorial plane. Furthermore we are only interested in prograde, circular\footnote{Circular here means $r=\text{const}.$ in the Boyer-Lindquist type coordinates; the shape of the orbit is not necessarily a geometric circle.} orbit.
We emphasize that there is a subtlety regarding what it \emph{means} to have a photon orbit on the horizon for a rotating black hole. This problem was discussed
in the context of an asymptotically flat Kerr black hole, of which
it is well-known that although photon orbit and the horizon do coincide in the Boyer-Lindquist coordinates, this is simply because the coordinates become degenerate in the extremal limit. This can be clearly seen in the embedding diagram (see, e.g., Fig.(2) in \cite{BPT}), in which the black hole throat tends to a cylinder in the extremal limit. Therefore, despite being two separate circles on this cylinder, both the photon orbit and the horizon have the same radius for an extremal black hole. Nevertheless, Jacobson \cite{1107.5081} showed that in the Doran frame \cite{9910099}, which is analogous to the Painlev\'e-Gullstrand frame for the Schwarzschild black hole, the photon orbit \emph{does} coincide with the horizon. Thus, in this work, when we refer to photon orbit on the horizon of an extremal rotating black holes, it is to be understood that this statement should be interpreted in the Doran sense. If one prefers to think in terms of the Boyer-Lindquist coordinates, then we are concerned with the \emph{coordinate} equality $r_h=r_p$, where $r_h$ and $r_p$ shall denote throughout this work, respectively, the event horizon and the photon orbit. One might wonder if this is physically meaningful, since whether the horizon coincides with the photon orbit (either in the physical sense or coordinate sense) depends on the choice of spacetime slices, and in principle we might  derive different results based on other coordinates. We shall return to this issue shortly.

\begin{figure}[h!]
\centering
\includegraphics[width=4.0in]{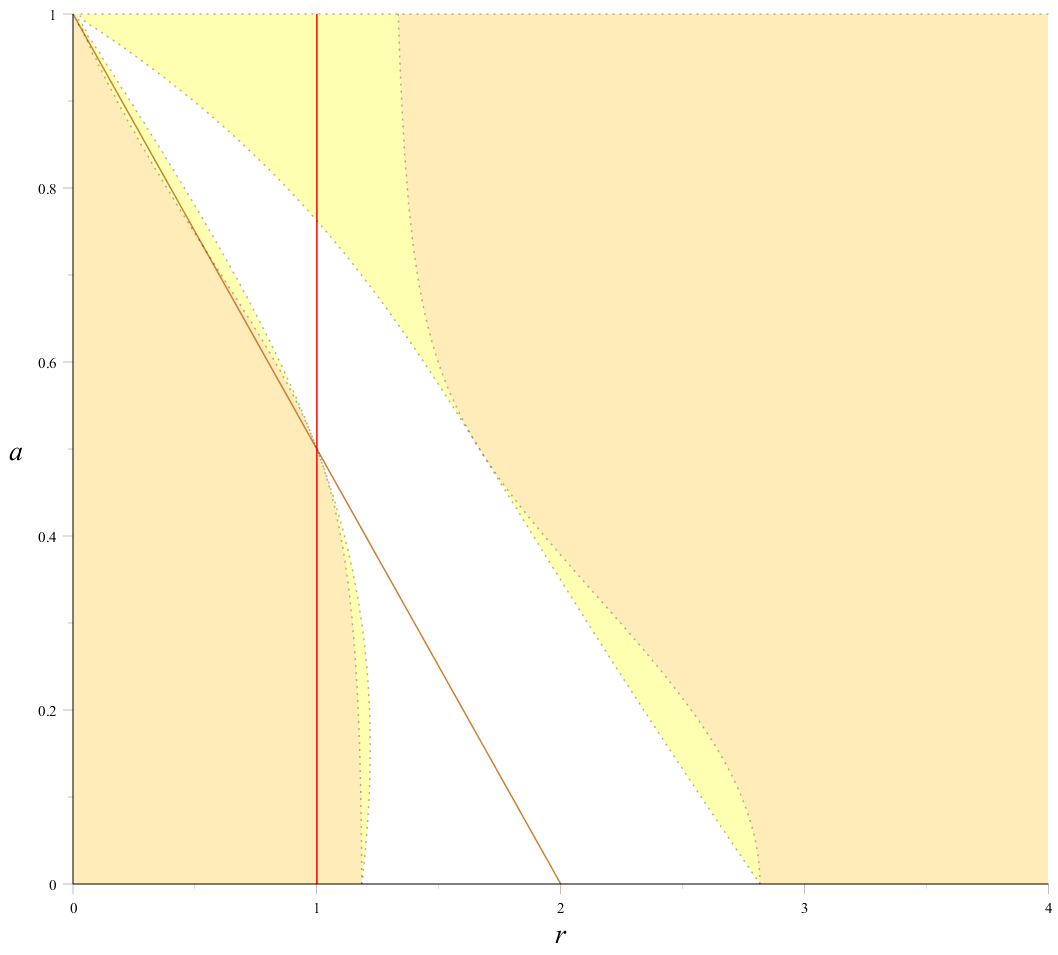}
\caption{This plot shows the location of the (prograde) photon orbits of an extremal asymptotically flat Kerr-Newman black hole. The solid vertical line coincides with the extremal horizon, which we set to be $r_h=1=M$. The common shaded regions (orange) are where both photon orbits are stable, whereas the other photon orbit, which is not on the horizon, has larger stability regions (shaded in yellow). For $a>M/2$, there is an unstable photon orbit on the horizon. However, for $a<M/2$, the horizon photon orbit stabilizes. At the same time, there exists another unstable prograde orbit outside the black hole.
See \cite{Khoo:2016xqv} for details. \label{regionplot}}
\end{figure}

\begin{figure}[h!]
\centering
\includegraphics[width=4.0in]{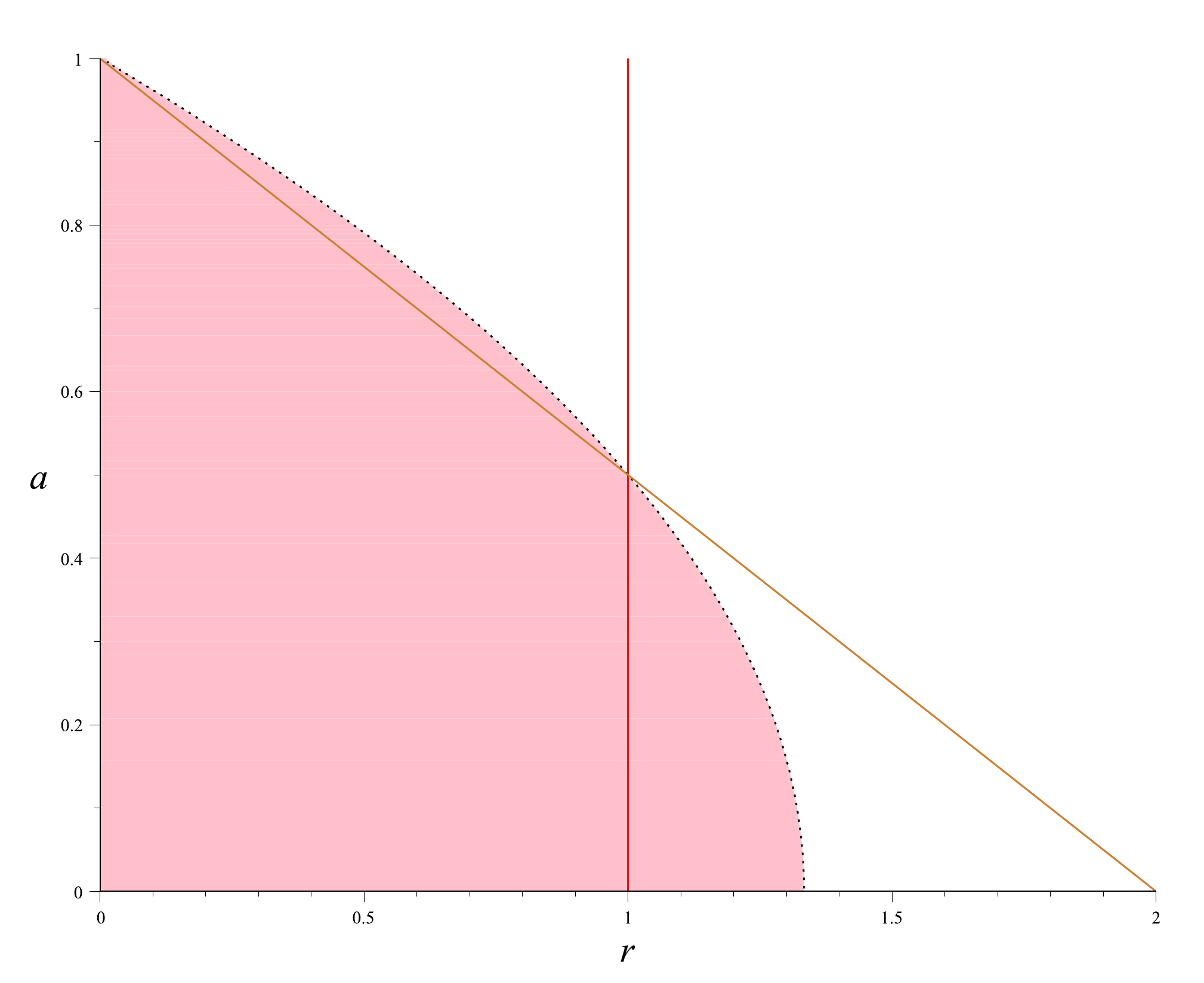}
\caption{The same photon orbits of an extremal Kerr-Newman black hole as plotted in Fig.(\ref{regionplot}). Here, the shaded region is the stability region obtained from  $V''_\text{eff}(b,r)>0$, where $b$ is the root of $V_\text{eff}'(b,r)=0$. See main text for more details. \label{plotaf}}
\end{figure}

The stability of the photon orbit on the horizon of an extremal asymptotically flat Kerr-Newman black hole was discussed recently by Ulbricht and Meinel \cite{1503.01973}, in which they found that the photon orbit only coincides with the extremal horizon if the angular momentum is sufficiently large: $|J| > M^2/2$. On first look this seems to be inconsistent with the result of Pradhan and Majumdar \cite{1001.0359}, which showed that there is a stable photon orbit on the event horizon of an extremal Reissner-Nordstr\"om black hole, and such a black hole is just a special case of the Kerr-Newman family with $|J|=0 < M^2/2$. However, there is no real inconsistency, as clarified by Khoo and Ong \cite{Khoo:2016xqv}: what really happens in the Kerr-Newman case is that for a sufficiently large angular momentum $|J| > M^2/2$, there is in fact an unstable photon orbit on the extremal horizon. For $|J| < M^2/2$, there is also a photon orbit on the extremal horizon, but it is a \emph{stable} one, which indeed agrees with the result of Pradhan and Majumdar \cite{1001.0359}. See Fig.{\ref{regionplot}} below for an illustration. In this figure, to obtain the photon orbit one uses the conditions on the effective potential\footnote{The effective potential of massless particles defined herein satisfies in general $V_\text{eff}+\dot{r}^2/J_0^2=1/b^2$, where $b$ plays the role of an impact parameter in the asymptotically flat cases, and $J_0$ is the angular momentum of the particle.} $V_\text{eff}(b,r)=1/b^2$, and $V'_{\text{eff}}(b,r)=0$. 
For each photon orbit, there corresponds a fixed $b$. One then studies the effective potential $V_\text{eff}(b,r)$, the stability of which is governed by the sign of the second derivative $V''_\text{eff}(b,r)$. This was the method used in \cite{Khoo:2016xqv}. Alternatively, one could differentiate $V_\text{eff}(b,r)$ twice with respect to $r$, while holding $b=\text{const.}$, and then study the sign of $V''_\text{eff}(b,r)$, where $b$ is the root of $V_\text{eff}'(b,r)=0$. The stability region obtained is shown in Fig.{\ref{plotaf}}. In so far as the photon orbits are concerned, they lie in the stability region obtained using in the first method if and only if they lie in the stability region obtained using the second method. The second method is nevertheless cleaner. 

The investigation above concerning an extremal asymptotically flat Kerr-Newman black hole \cite{Khoo:2016xqv,1503.01973}
suggests that: angular momentum tends to destabilize the photon orbit on an extremal black hole horizon, whereas electrical charge tends to stabilize it. The question we are interested in is the following: \emph{How much does this behavior persist beyond this very specific example?} 

To this end, we decided to study stable photon orbits on extremal black hole horizons in asymptotically anti-de Sitter (AdS) spacetimes. This allows us to see whether the aforementioned behavior changes with the introduction of a negative cosmological constant, especially when rotation is involved. The behavior of particle orbits around black holes with nonzero cosmological constants had been studied rather extensively in the literature, see, e.g., \cite{0307049,0803.2539,1267280,4500311}. However, our objective here is much more specific and the question raised in the previous paragraph does deserve a more focused and cleaner analysis. 

To further motivate our choice to focus on asymptotically AdS spacetimes, it is well-known that black holes in such spacetimes play very important roles in the context of AdS/CFT (Conformal Field Theory) correspondence (or more generally, holography) \cite{maldacena,gibbons}, and so any further understanding we gain from studying the properties of these black holes may, via the holographic correspondence, lead to new understanding of the dual field theory. In this work, we shall focus on two examples: BTZ black holes in $(2+1)$-dimensions -- which also allows us to explore the effect of lower spacetime dimensions on the stability of photon orbit -- and Kerr-Newman-AdS black holes in $(3+1)$-dimensions. 

Furthermore, in the context of holography, the standard Schwarzschild-type coordinates (or the Poincar\'e-type coordinate in which $z$ is simply inversely proportional to the radial coordinate $r$) for static, asymptotically locally AdS black holes, actually do have \textit{physical} importance. Recall that the Hawking radiation from a black hole also depends on the observer. Notably, the standard Hawking temperature in the asymptotically flat case corresponds to the temperature measured by an observer infinitely far away from the black hole. A different slicing of the spacetime would yield a different temperature. In holography, the black hole temperature corresponds to the temperature of the dual field theory, and it is precisely in these coordinates that Hawking temperature is typically calculated in the bulk. Similarly, in the case of rotating black holes in the AdS bulk, one uses the standard, Boyer-Lindquist-type coordinates, to compute the Hawking temperature, which is then used to study some dual field theoy with nonzero angular momentum. Thus, arguably, Boyer-Lindquist-type coordinates do have straightforward, physical importance in holography. This justifies why we work in these coordinates, instead of some others, given that whether the photon orbit coincides with the extremal horizon depends on how one slices the spacetime. Of course, the implication of the physics of photon orbit for the dual field theory, if any, remains to be studied but is beyond the scope of our work.

We will be working in the units in which $c=G=1$, although sometimes we do restore $G$ for the sake of clarity, or for emphasis.

\section{Photon Orbits on the Horizons of Extremal $\text{AdS}_{3}$ Black Holes}\label{3}

It is a curious but well-known fact that general relativity without the cosmological constant term in $(2+1)$-dimensions is trivial: the vacuum Einstein equations $R_{\mu\nu}=0$ implies that the full Riemann tensor vanishes identically: $R_{\mu\nu\lambda\rho}\equiv 0$. In particular, this means that there is no black hole solution. 
It therefore came at a pleasant surprise that Einstein's gravity in $(2+1)$-dimensions is far richer if one introduces a negative cosmological constant. This entirely new field of research in lower dimensional gravity began with the famous black hole solution by Ba\~{n}ados, Teitelboim (now Bunster) and Zanelli \cite{BTZ1, BTZ2} -- hereinafter, the BTZ black hole. It was subsequently realized that a large variety of spacetimes, which includes the BTZ black holes, can be obtained by identifying points in (2+1)-dimensional AdS spacetime by means of a discrete group of isometries \cite{9707036}. In fact, the Euclidean version of the BTZ manifold is simply $\Bbb{H}^3/\Gamma$, where $\Gamma \subset \text{PSL}(2,\Bbb{C})$ is a Schottky group \cite{005106}.

Since their discoveries, BTZ black hole and its variations have been widely studied and yielded many surprises in the past several decades. Due to its simplicity, many toy models in $(2+1)$-dimensions are used to help us gain insights into the corresponding physics in higher dimensions.  
A recent example is a toy model of the Riemannian Penrose inequality, which has been proved in $(2+1)$-dimensions in AdS background \cite{1608.06092} (there is still no general proof for the inequality in $(3+1)$-dimensions). BTZ black holes have also found some surprising applications, e.g., in modeling the physics of graphene \cite{1202.2938, 1205.4039}, via the optical metric approach. 

Since the BTZ solutions are asymptotically AdS, many studies regarding AdS/CFT correspondence have been performed on such background spacetimes, see \cite{Carlip, 9808037, 1307.7738}. Indeed, since general relativity becomes a topological field theory in three spacetime dimensions (i.e., it has no propagating gravitational degrees of freedom), its dynamics can be largely described holographically by a two-dimensional CFT at the boundary of the spacetime \cite{Carlip}. Therefore, the BTZ black hole provides an excellent example of how AdS/CFT correspondence works, albeit in this very special example. 
 
In the section, we will study the photon orbits of BTZ black holes. Specifically, we would like to see if in the extremal case there are photon orbits on the horizon, and if so whether they are stable.  Since there is no $(2+1)$-dimensional black hole in the absence of cosmological constant, the results in this section have no asymptotically flat cousins to compare with, but it will be an important study on whether a lower dimensional spacetime affects the behavior of the photon orbits differently.

\subsection{Case I: Extremal Static Charged BTZ Black Hole}\label{2.1}

We begin with the simplest case: the charged but non-rotating BTZ black hole. 
The spacetime is therefore static. The proof in \cite{Khoo:2016xqv} concerning the stability of photon orbit on the horizon of static extremal black holes in fact
also works in $(2+1)$-dimensions, and so we know without explicit calculation that for this case, there is also a stable photon orbit on the extremal horizon. However, since
this spacetime has many interesting properties, we feel that there is a need for a more explicit treatment.  

The metric tensor of a charged but non-rotating BTZ black hole is
\begin{equation}
\d s^{2}=-f(r)\d t^{2}+\frac{\d r^{2}}{f(r)}+r^{2}\d\theta ^{2},
\end{equation}%
in which 
\begin{equation}\label{BTZ}
f(r)=\frac{r^{2}}{\ell^{2}}-m-2q^{2}\ln \left({\frac{r}{\ell}}\right),
\end{equation}
where $m$ and $q$ are, respectively, the mass and the charge parameters of the black hole,
and $\ell$ is the asymptotic curvature radius of the spacetime. The physical mass, $\mathcal{M}$, and physical charge, $\mathcal{Q}$, differ from $m$ and $q$ only by overall constants \cite{1507.08496}:
\begin{equation}
\mathcal{M}=m/8, ~~\mathcal{Q}=q/2.
\end{equation} 
Note that the metric coefficient in Eq.(\ref{BTZ}) contains the asymptotic curvature length scale $\ell$ in the logarithmic term: this is important to ensure that the argument of logarithm is dimensionless. This length scale could in principle be free, i.e., \emph{any} length scale could be introduced for this purpose, and the field equations will be satisfied. Nevertheless, $\ell$ is a natural length scale of the system. In addition, it plays an important role in removing the ensemble dependency in the thermodynamics of the system \cite{1507.08496}. 

Solving the equation $f(r)=0$, we can obtain the location of the horizon, $r_{h}$. 
Furthermore, if the black hole is extremal, then its horizon would also satisfy $f'(r)=0.$
We shall denote the extremal horizon by $r_{c}$, where the subscript ``$c$'' refers to ``critical''. The extremal horizon satisfies

\begin{equation}
f(r_{c})=\frac{r_{c}^{2}}{\ell^{2}}-m_{c
}-2q_{c}^{2}\ln\left( {\frac{r_{c}}{\ell}}\right)=0,\text{ \ and \ }
f^{\prime }(r_{c})=\frac{2r_{c}}{\ell^{2}}-\frac{2q_{
c}^{2}}{r_{c}}=0.
\end{equation}
The extremal horizon is therefore $r_{c}=q_{c}\ell$. 
Note that we now have an equation that relates $q_{c}$ and $m_{c}$. This is the extremal condition (see also \cite{1610.01744}):
 
\begin{equation}
q_{c}^{2}-2q_{c}^{2}\ln{q_{c}}=m_{c}.  \label{eq:q}
\end{equation}

\begin{figure}[h]
\centering
\includegraphics[width=0.65\textwidth]{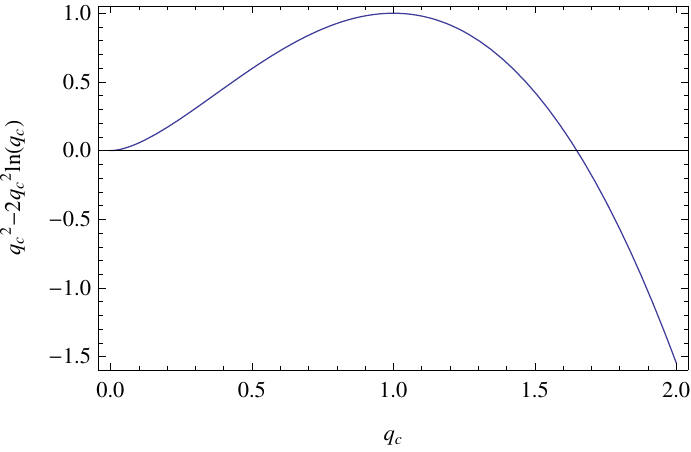}
\caption{{}The plot of the left hand side expression of Eq.(\protect\ref{eq:q})
against $q_{c}$. There is a global maximum at $q_{c}=1$. }
\label{fig:q}
\end{figure}

The behavior of the left hand side expression of Eq.(\ref{eq:q}) is shown in Fig.\ref{fig:q}. It has a peak of unity when $q=1$. 
This means that, for mass $m=1$, there is only one solution for Eq.(\ref{eq:q}), which is at $q=1=q_{\text{ext}}$, and the charged BTZ black hole is extremal.  
When $0<m<1$, there will be two solutions for $q_{\mathrm{ext}}$ from Eq.(\ref{eq:q}). The black hole can be extremal for these two values of the charge when $0<m<1$. However, when $m>1$, there is no solution to Eq.(\ref{eq:q}). This indicates that the extremal black hole condition cannot be satisfied. In other words, charged BTZ black holes are \emph{always} non-extremal when $m>1$. 

Note that for a sufficiently large charge (which can be arbitrarily large), negative mass black hole solutions exist. See also the discussion in \cite{9912259}.  For this work, we will focus on black holes with positive mass. Indeed, the notions of mass and charge (and angular momentum) are rather peculiar in $(2+1)$-dimensions, and one finds in the literature various attempts to make sense of what counts as the \emph{physical} parameters of a black hole \cite{0301129,0905.3517}. For example, 
a different notion of black hole mass, $M_0$, which is essentially the energy contained inside the horizon, was proposed in \cite{0905.3517}. It is given by a combination of $m$ and $q$, specifically $M_0(r_h)=m + 2q^2\ln (r_h/\ell)$. Such a mass does provide a BPS-like bound. For our purpose however, we need not worry about these subtleties, since \emph{the extremal condition is unambiguously known}, and this condition can be expressed using any variables one wishes without changing the underlying physics. (This does not mean that identifying the correct physical quantities is not important in other contexts, as we will remark in Sec.\ref{3.1}.)

Since charged BTZ black holes are static, we can easily calculate the effective potential $V_{\text{eff}}$ experienced by a massless particle:

\begin{equation}
V_\text{eff}(r)=\frac{1}{r^{2}}f(r).
\end{equation}

The location $r_{p}$ of a photon orbit corresponds to an extremal point of the potential, whose stability depends on whether $V''_\text{eff}(r)$ is positive or negative. The explicit expressions for $V_\text{eff}(r)$, $V'_\text{eff}(r)$ and $V''_\text{eff}(r)$ are as follows:

\begin{equation}
\begin{cases}
V_\text{eff}(r)=\dfrac{1}{r^{2}}f(r)=\dfrac{1}{r^{2}}\left[\dfrac{r^{2}}{\ell^{2}}-m-2q^{2}\ln \left({\dfrac{r}{\ell}}\right)\right],\\
\\
V'_\text{eff}(r)=\dfrac{1}{r^{2}}\left[f'(r)-\dfrac{2}{r}f(r)\right]=\dfrac{2}{r^{3}}\left[m-q^{2}+2q^{2}\ln\left({\dfrac{r}{\ell}}\right)\right],\\
\\
V''_\text{eff}(r)=\dfrac{1}{r^{2}}\left[f''(r)-\dfrac{4}{r}f'(r)+\dfrac{6}{r^{2}}f(r)\right]=\dfrac{2}{r^{4}}\left[5q^{2}-3m-6q^{2}\ln\left({\dfrac{r}{\ell}}\right)\right].
\end{cases}
\end{equation}

As we have explained, for $m>1$, there is no extremal black hole. For $m=1$, there is only one extremal charge $q_{c}=1$ that can satisfy the extremal condition. In this case, the extremal horizon is at $r_{c}=\ell $, while the extremal point of the potential, $r_{p}$, is also equal to $\ell$. Furthermore $V''_\text{eff}(\ell)\geqslant 0$, so $r_{p}=r_{h}=\ell$ is a local minimum. This corresponds to the only stable photon orbit of this spacetime. For $0<m<1$, there are two extremal charges $q_{c_1}$ and $q_{c_2}$, which corresponds to the same value of the mass. When $q_{c}=q_{c_1}$, we have
\begin{equation}
q_{c_1}^{2}-2q_{c_1}^{2}\ln{q_{c_1}}=m_{c},
\end{equation}
so
\begin{equation}
V'_\text{eff}(r)=\frac{2}{r^{3}}\left[m_{c}-q_{c_1}^{2}+2q_{c_1}^{2}\ln\left(\frac{r}{\ell}\right)\right]=\frac{2}{r^{3}}\left[-2q_{c_1}^{2}\ln{q_{c_1}}+2q_{c_1}^{2}\ln\left(\frac{r}{\ell}\right)\right].
\end{equation}
We set $V'_\text{eff}(r)=0$ to obtain $r_{p_1}=q_{c_1}\ell=r_{c_1}$.  Since local minimum requires that
\begin{equation}
V''_\text{eff}(r)=\frac{2}{r^{4}}\left[5q_{c_1}^{2}-3m_{c}-6q_{c_1}^{2}\ln\left(\frac{r}{\ell}\right)\right]=\frac{4q_{c_1}^{2}}{r^{4}}\left[1+3\left(\ln{q_{c_1}}-\ln\left(\frac{r}{\ell}\right)\right) \right]\geqslant 0,
\end{equation}
stable photon orbits need $r_{p_1}\leqslant {e^{1/3}} q_{c_1}\ell \approx 1.396 q_{c_1}\ell$, which is obviously satisfied. 

In the same manner, we can verify that when  $q_{c}=q_{c_2}$, we have similarly $r_{p_2}=q_{c_2}\ell=r_{c_2}$, which is also a stable photon orbit.

The figure of the photon orbits and the stable regions are given as follows.

\begin{figure}[h!]
\centering
\includegraphics[width=0.45\textwidth]{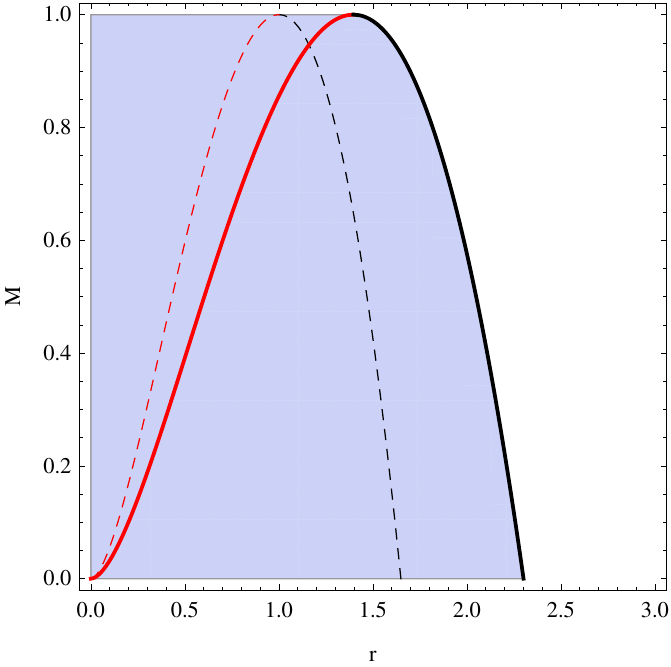}
\caption{The red dashed curve and the black dashed curve, shown in a $M$-versus-$r$ plot, correspond to photon orbits $r_{p_1}$ and $r_{p_2}$, respectively. The shaded region to the left of the solid red curve corresponds to the stable region of photon orbit $r_{p_1}$, while the shaded region to the left of the solid black curve corresponds to the stable region of photon orbit $r_{p_2}$. }
\label{fig:Crp}
\end{figure}

\begin{table}[h!]
\caption{Photon Orbits for Static Charged BTZ Black Holes (BHs)}
\centering
\begin{tabular}{|c|c|c|}
\hline
\multicolumn{3}{|c|}{Charged BTZ Black Holes} \\ 
\hline
$m>1$ & $m=1$ & $0<m<1$\\
\hline
No extremal BHs & One extremal case when $q=1$ & Two extremal cases when $q=q_{1}$ or $q=q_{2}$\\
\hline
\ & $r_{p}=r_{h}=\ell$ & $r_{p_1}=r_{h_1}=q_{1}\ell$ or $r_{p_2}=r_{h_2}=q_{2}\ell$\\
\hline
\ & Stable & Stable\\
\hline
\end{tabular}
\end{table}

To conclude, for an extremal static charged BTZ black hole, there always exists a stable photon orbit on its horizon.
This result is the same as the case of asymptotically flat Reissner-Nordstr\"om black holes in $(3+1)$-dimensions. This is as expected, since as we mentioned at the beginning of Sec.\ref{3}, the charged BTZ solution is subjected to the theorem proved in \cite{Khoo:2016xqv}\footnote{Although  \cite{Khoo:2016xqv} focused on the asymptotically flat case, said theorem is applicable also in asymptotically AdS case. (As mentioned in that work, the case for asymptotically de-Sitter black holes is more subtle and the theorem does not apply.) The crucial assumptions of the theorem are spherical symmetry of the spacetime and analyticity of the metric. It is only when we deviate away from spherical symmetry that the effect of the cosmological constant becomes important.}.

\subsection{Case II: Extremal Rotating BTZ Black Hole}

We now consider an uncharged rotating BTZ black hole, whose metric tensor is 
\begin{equation}
\d s^{2}=-N^{2}\d t^{2}+N^{-2}\d r^{2}+r^{2}\left(N^{\phi} \d t+\d\phi\right)^{2},
\end{equation}
where
\begin{equation}
N^{2}=-m+\frac{r^{2}}{\ell^{2}}+\frac{J^{2}}{4r^{2}}, \ \ \text{and} \ \  N^{\phi}=-\frac{J}{2r^{2}},
\end{equation}
in which $J$ denotes the angular momentum of the black hole.

We  write
\begin{equation}
g(r) \equiv N^{2}=-m+\frac{r^{2}}{\ell^{2}}+\frac{J^{2}}{4r^{2}}, 
\end{equation}
which gives us
\begin{equation}
g'(r)=\frac{\partial N^{2}}{\partial r}=\frac{2r}{\ell^{2}}-\frac{J^{2}}{2r^{3}}.
\end{equation}
The event horizon is located at
\begin{equation}
r_{h}=\ell m^{\frac{1}{2}}\left[\frac{1\pm \sqrt{1-\left(\frac{J}{m\ell}\right)^{2}}}{2}\right]^{\frac{1}{2}},
\end{equation}
where $m>0$ and $ |J | \leqslant m\ell$.

We now consider the extremal case. Setting both the metric coefficient $g(r)$ and its derivative to be zero, we obtain

\begin{equation}
J_{c}=m_{c}\ell, \ \ \ \ \ \  r_{c}=\sqrt{\frac{J_{c}\ell}{2}}=\ell\sqrt{\frac{m_{c}}{2}}.
\end{equation}

Following the usual procedure\footnote{There is, however, a subtlety due to the rotating nature of the spacetime: the ``effective potential'' is $b$-dependence. See also \cite{hobson}. Note that the definition of the effective potential could differ by either an additive constant or an overall constant in different texts.} \cite{Robert:1984, Derek:2015}, we can calculate the effective potential $V_{\text{eff}}$ experienced by a massless particle as follows,
\begin{equation}
V_\text{eff}(r)=\frac{N^{2}}{r^{2}}-\frac{2N^{\phi}}{b}-N^{2\phi}=\frac{1}{r^{2}}\left(\frac{r^{2}}{\ell^{2}}-m+\frac{J}{b}\right),
\end{equation} 
in which $b={J_{0}}/{E_{0}}$, where $J_{0}$ and $E_{0}$ denote, respectively, the angular momentum and the energy of the particle. For an asymptotically flat spacetime, $b$ can be interpreted as an impact parameter, however such an interpretation does not quite exist in an asymptotically AdS spacetime. 

In the extremal case, the effective potential and its derivative are
\begin{equation}
V_\text{eff}(r)=-\frac{1}{r^{2}}\left(m_{c}-\frac{m_{c}\ell}{b}-\frac{r^{2}}{\ell^{2}}\right), \ \ \ \ \ V_\text{eff}'(r)=-\frac{2m_{c}}{r^{3}}\left(\frac{\ell}{b}-1\right).
\end{equation}

Obviously there is only one solution for $V_\text{eff}'(r)=0$, given by $b={J_{0}}/{E_{0}}=\ell$. In this case the effective potential is always a constant $V_\text{eff}=\ell^{-2}=b^{-2}$. This result agrees with \cite{9401025}.

Thus, for the case of a rotating but uncharged extremal BTZ black hole,  we can only have photon orbits when $b=\ell$. These photon orbits are all ``borderline'' unstable, in the sense that a photon on a given photon orbit $r_{p^*}$, can, under a perturbation (say, if scattered by other particles), end up on a neighboring photon orbit $r_{p^*+\varepsilon}$, for any $\varepsilon > 0$. In fact, a constant potential means that a particle can easily roll across it, and therefore is extremely unlikely to remain in any fixed orbit for long.




\subsection{Case III: Charged Rotating BTZ Black Hole} 

Finally, let us consider the most general case: a BTZ black hole with both charge and angular momentum. Here we only study the effect of electrical charge, though the magnetic counterpart is known \cite{0201058}. 
The metric tensor for a rotating, (electrically) charged black hole\footnote{Note that the correct metric for a charged and rotating BTZ black hole is \emph{not} given by a straightforward generalization by adding a rotation term in the charged BTZ solution (such as those analyzed in \cite{0912.0861,1009.3749}, among many other works), as clarified long ago in the literature \cite{9505037,9909111}. We restore the Newton's constant $G$ in this section for clarity since this metric does not seem to be as well known as it should be.}, is given by\cite{9912259,9510025} 

\begin{equation}
\d s^{2}= -N^{2}\d t^{2}+K^{2}\left(\d\phi+N^{\phi}\d t\right)^{2}+\frac{r^{2}\d r^{2}}{K^{2}N^{2}}, 
\end{equation}
in which 
\begin{equation}
N^{2}=\frac{r^{2}}{K^{2}}\left(\frac{r^{2}}{\ell^{2}}-\frac{\overline{\ell}^{2}}{\ell^{2}}8\pi G Q^{2}\ln{\frac{r}{\overline{r}_{0}}}\right),\ 
N^{\phi}=-\frac{\omega}{K^{2}}8\pi G Q^{2}\ln{\frac{r}{\overline{r}_{0}}},
\end{equation}
\begin{equation}
K^{2}=r^{2}+\omega^{2}8\pi G Q^{2}\ln{\frac{r}{\overline{r}_{0}}},\ 
A_{\mu}\d x^{\mu}=Q\ln{\frac{r}{\overline{r}_{0}}}\left(\d t-\omega \d\phi\right),
\end{equation}
\begin{equation}
\overline{r}_{0}=\frac{\overline{\ell}}{\ell}r_{0},\ 
\overline{\ell}^{2}=\ell^{2}-\omega^{2}\ \ \left(|\omega|<\ell\right).
\end{equation}
Here $A_\mu \d x^\mu$ is the usual electromagnetic gauge potential, $\omega$ is the angular velocity of the black hole.
The charge $Q$ here is related to the charge parameter $q$ in subsect.(\ref{2.1}) by $Q^2=q^2/4\pi$. (That is, $q$ is the electric charge parameter in the Lorentz-Heaviside units.) Again, there are non-trivial subtleties about whether $Q$ is the \emph{physical} charge \cite{9912259}, but as remarked in Sec.\ref{2.1}, this does not affect our analysis.

Note that, following \cite{9510025}, we do not include an explicit mass term, $m$, in the metric. The mass term depends on the choice of $r_0$ (which is left unfixed in \cite{9510025}), that is, the set of black hole hairs $\left\{m,Q,J\right\}$, can be traded with the set of parameters $\left\{r_0, Q,\omega\right\}$. (See also \cite{0206024}.)
However, if we were to recover the metric of the form of Eq.(\ref{BTZ}) in the $\omega \to 0$ limit, then our $r_0$ is in fact \emph{fixed}, it is
\begin{equation}
r_0 = \ell \exp\left(-\frac{m}{8\pi Q^2}\right).
\end{equation}
That is, $r_0$ depends on the values of $m$ and $Q$.

The extremal horizon and extremal charge are, respectively,
\begin{equation}
r_{c}=\overline{r}_{0_c}e^{\frac{1}{2}},\ Q_{c}=\frac{r_{0_c}e^{\frac{1}{2}}}{2\ell\sqrt{\pi G}}.
\end{equation}
Note that the extremal charge depends on $r_{0_c}$, which itself contains $m_c$ and $Q_c$. Explicitly, 
\begin{equation}
m_c  = 4\pi Q_c^2\left[1-2\ln\left(2\sqrt{\pi G}Q_c\right)\right].
\end{equation}

We can then calculate the effective potential for massless particle and its derivative. They are

\begin{equation}
\begin{cases}
V_\text{eff}(r)=\dfrac{1}{r^{2}}\left(N^{2}-K^{2}N^{2\phi}-\dfrac{K^{2}}{b^{2}}-\dfrac{2K^{2}N^{\phi}}{b}\right)+\dfrac{1}{b^{2}}
\\
\\
\ \ \ \ \ \ \ \ \ =\dfrac{1}{b^{2}\ell^{2}r^{2}}\left[b^{2}r^{2}-8\pi G \ell^{2}Q_{c}^{2}\left(b-\omega\right)^{2}\ln{\dfrac{r}{\overline{r}_{0_c}}} \right],
\\
\\
V_\text{eff}'(r)=\dfrac{2}{b^{2}\ell^{2}r^{3}}\left[\left(\ell^{2}-b^{2}\right)r^{2}+8\pi G \ell^{2}Q_{c}^{2}\left(b-\omega\right)^{2}\ln{\dfrac{r}{\overline{r}_{0_c}}} \right].
\end{cases}\label{Veff}
\end{equation}

Note that if $b=\omega$, then we see that the second term vanishes in the equation Eq.(\ref{Veff}), and $b^{2}r^{2}$ cancels in the remaining term, so that $V_\text{eff}={1}/{\ell^{2}}$. Therefore had it was possible for $b=\ell=\omega$, we would have a photon orbit. The reason we do not is because the condition $|\omega| < l$ forbids this case. If $b=\ell\neq \omega$, then
\begin{equation}
V_\text{eff}'(r)=\dfrac{2}{\ell^{4}r^{3}}\left[8\pi G \ell^{2}Q_{c}^{2}\left(b-\omega\right)^{2}\ln{\frac{r}{\overline{r}_{0_c}}} \right], 
\end{equation}
we can see that $r=\overline{r}_{0_c}$ is an unstable photon orbit $\left(V_\text{eff}''\left(\overline{r}_{0_c}\right)=0\right)$, which is inside the horizon.

When $b\neq\omega$ and $b\neq\ell$, if we want $r_{p}=r_{c}$, $b$ must satisfy the relation
\begin{equation}
b=\frac{\ell^2}{\omega}, ~~\omega \neq 0\label{brelation}
\end{equation}

\begin{figure}[h]
\centering%
\includegraphics[width=.45\textwidth]{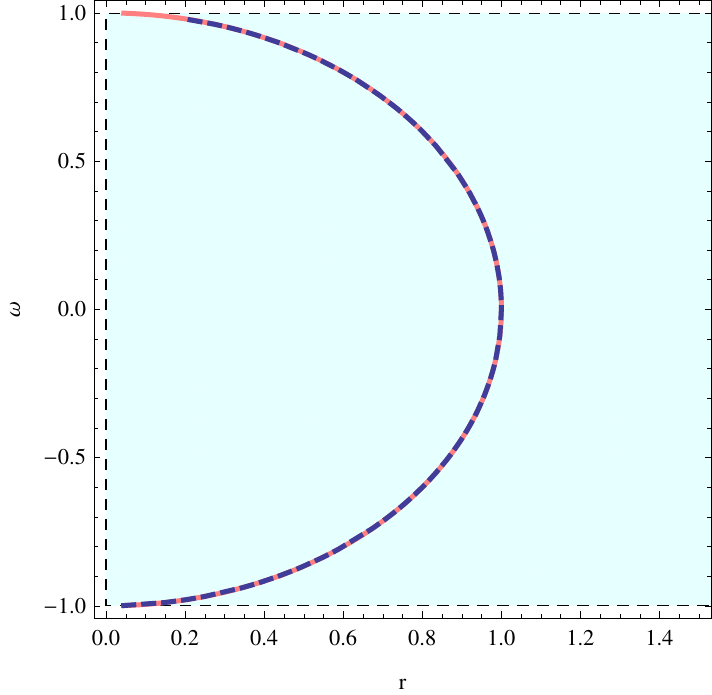} 
\caption{{} This plot shows that the photon orbit (the blue dashed curve) coincides with the horizon (the red curve) for an extremal charged rotating BTZ black holes with $\ell =m=1$ and the units chosen such that $G\pi=1$. The photon orbits in the shaded region are stable. Note that here we set $b={\ell^2}/{\omega}$. }
\label{CR}
\end{figure}

In Fig.\ref{CR}, we plotted the location of the photon orbit and the event horizon of an extremal charged rotating BTZ black hole with $\ell =m=1$, and with $b$ satisfying the relation in Eq.(\ref{brelation}) so that the photon orbit coincides with the event horizon. To be consistent with the previous cases, here we have set $\pi G=1$.

From Fig.\ref{CR}, we can see that there is a stable photon orbit on the horizon of an extremal charged rotating BTZ black hole.

\section{Photon Orbits on the Horizons of Extremal $\text{AdS}_4$ Black Holes}

Having completed our analysis of photon orbits on the extremal horizons of BTZ black holes, we now move on to asymptotically AdS black holes in $(3+1)$-dimensions. We consider only black holes whose horizon is topologically a sphere, so that the results in this section can be compared directly with the results in \cite{Khoo:2016xqv}, which concerns the asymptotically flat case. Our strategy is to start with the most general case with both rotation and charge, and then specialize to the purely rotating case. The extremal charged case without rotation (i.e. Reissner-Nordstr\"om-AdS black holes) will not be discussed here since it was already shown in \cite{Khoo:2016xqv} that such a black hole possesses a stable photon orbit on its event horizon. See also \cite{0803.2685}.

\subsection{Case I: Kerr-Newman-AdS Black Hole} \label{3.1}

The 4-dimensional Kerr-Newman-AdS solution describes a rotating and charged black hole in $\text{AdS}_4$. 
This is an important family of black holes in the context of holography, with applications in, e.g., the holography of quark-gluon plasma produced in heavy ion collisions \cite{1403.3258}.

The metric tensor of this black hole can be written in the Boyer-Lindquist form

\begin{equation}
\d s^{2}=-\frac{\Delta_{r}}{\Xi^{2}\rho^{2}}\left[\d t-a\sin^{2}{\theta}\d\phi \right]^{2}+\frac{\rho^{2}}{\Delta_{r}}\d r^{2}+\frac{\rho^{2}}{\Delta_{\theta}}\d\theta^{2}+\frac{\Delta_{\theta}}{\Xi^{2}\rho^{2}}\sin^{2}{\theta}\left[a\d t-\left(a^{2}+r^{2}\right)\d\phi \right]^{2},
\end{equation}
in which
\begin{equation}
\left\{
\begin{array}{ll}
\rho^{2}=r^{2}+a^{2}\cos^{2}{\theta}, \\

\Delta_{r}=\left(a^{2}+r^{2}\right)\left(1+\frac{r^{2}}{\ell^{2}}\right)-2mr+e^{2}, \\

\Delta_{\theta}=1-\frac{a^{2}\cos^{2}{\theta}}{\ell^{2}},\\

\Xi=1-\frac{a^{2}}{\ell^{2}}.
\end{array}
\right.
\end{equation}
Here $m$, $e$, and $a$ are, respectively, the mass parameter, the charge parameter, and the rotation parameter. 
We will assume that $m,e,a > 0$. 
The physical (Abbott-Deser \cite{AD}) mass and physical charge of the black hole (i.e., the mass and charge that actually satisfy the laws of black hole thermodynamics \cite{0408217}, as well as being the correct conserved quantities associated to some Killing vectors \cite{0506057}) are, respectively, $m/\Xi^2$ and $e/\Xi$. Likewise, the physical angular momentum is $J=am/\Xi^2$. Note that the rotation parameter $a$ is the ratio of the \emph{physical} angular
momentum to the \emph{physical} mass. The distinction between a mere parameter and a physical quantity is quite important in considering various physics questions, such as in the attempt to ``over-spin'' a black hole, see \cite{1506.01248}. The parameter 
$a^*:=(a/m)(1-a^2/\ell^2)^2$ is useful in asymptotically flat case (as is usually used in astrophysics) to indicate how close to extremality the black hole is: $a^*=0$ indicates zero angular momentum, while $a^*\sim 1$ is near extremal. This is not a good indicator in asymptotically AdS spacetimes: even if $a^*$ is ``small'' (by asymptotically flat standard), say around 0.18, a black hole could nevertheless be spinning at 98.8\% of its maximal allowed speed \cite{1108.6234}, i.e., it is actually near extremal.

Note that for the metric to make sense, neither $\Xi$ nor $\Delta_{\theta}$ can vanish. This requirement imposes the strict inequality $a<\ell$, which means that black holes cannot rotate too fast in asymptotically AdS spacetimes. This upper limit on $a$ is independent of the more familiar, cosmic censorship requirement (see below). The simple inequality $a < \ell$ is in fact quite important, for mathematically, the function $\Delta_{\theta}$ has no fixed sign \textit{a priori} -- there is a risk that it can become negative near the poles. This would induce a change in the metric signature from $(-,+,+,+)$ to $(-,+,-,-)$ as one moves from the equator to the poles. The corresponding boundary field theory would then change from Lorentzian to Euclidean, which would be quite unphysical \cite{1403.3258, 1504.07344}. 

To study the photon orbit,
we set $\Delta_{r}$ and its derivative $\Delta_{r}':= {\d\Delta_{r}}/{\d r}$ to zero. 
This yields, for a fixed charge $e$,

\begin{equation}
\left\{
\begin{array}{ll}
r_{c}=\dfrac{1}{\sqrt{6}}\sqrt{-a^{2}-\ell^{2}+\sqrt{a^{4}+14a^{2}\ell^{2}+12e^{2}\ell^{2}+\ell^{4}}},\\
\\
m_{c}=\dfrac{a^{2}+\ell^{2}}{\ell^{2}}r_{c}+\dfrac{2}{\ell^{2}}r_{c}^{3},
\end{array}
\right.
\end{equation}
where $r_{c}$ is the extremal horizon, and $m_{c}$ is the corresponding extremal mass. The requirement for the existence of an event horizon (i.e. the cosmic censorship requirement) is $m\geqslant m_{c}$ or equivalently \cite{9808097},
\begin{flalign}
\frac{m}{\ell}\geqslant \Gamma\left(\frac{a}{\ell},\frac{e}{\ell}\right):=
&\frac{1}{3\sqrt{6}}\left[\sqrt{1+\frac{14a^{2}}{\ell^{2}}+\frac{12e^{2}}{\ell^{2}}+\frac{a^{4}}{\ell^{4}}}+\frac{2a^{2}}{\ell^{2}}+2\right] \\&\times \left[\sqrt{1+\frac{14a^{2}}{\ell^{2}}+\frac{12e^{2}}{\ell^{2}}+\frac{a^{4}}{\ell^{4}}}-\frac{a^{2}}{\ell^{2}}-1\right]^{\frac{1}{2}}.
\end{flalign}
See \cite{1403.3258,1506.01248} for more discussions.

The effective potential for a massless particle is given by
\begin{equation}
V_\text{eff}(r,b)=-\frac{\Xi^{2}}{b^{2}r^{4}}\left[\left(r^{2}+a^{2}-ab\right)^{2}-\left(a-b\right)^{2}\Delta_{r}\right]+\frac{1}{b^{2}},
\end{equation}
and its derivative is

\begin{equation}
V_\text{eff}'(r,b) =\frac{\partial V_\text{eff}(r,b)}{\partial r}=\frac{2(a-b)^{2}\left(\ell^{2}-a^{2}\right)^{2}}{b^{2}\ell^{6}r^{5}}\left[\frac{\left(a\ell^{2}+b\ell^{2}-a^{3}+a^{2}b\right)}{a-b}r^{2}+3m\ell^{2}r-2e^{2}\ell^{2}\right].
\end{equation}

\begin{figure}[h!]
\centering
\includegraphics[width=0.60\textwidth]{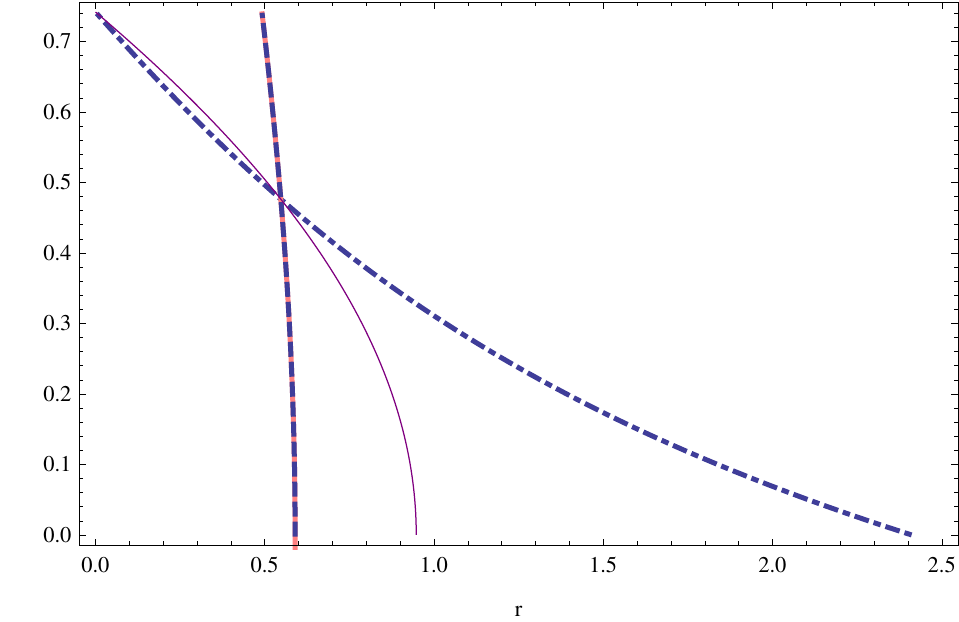}
\caption{{This plot shows the (prograde) photon orbits 
of an extremal asymptotically AdS Kerr-Newman black hole, here with $\ell =m=1$. There are two orbits denoted by the dashed blue curves, which correspond to two different values of ``impact parameter'' $b$. One of them coincide with the extremal horizon (thick line in red). The thin violet curve is the boundary of stability: any photon orbit to the left of it is stable. This should be compared to Fig.(\ref{plotaf}) for the asymptotically flat case.} \label{regionplot2} }
\end{figure}

The circular photon orbits can be obtained by solving simultaneously for the pair $\left\{b,r\right\}$ in the equations $V_\text{eff}(b,r)=1/b^2$ and $V_\text{eff}'(b,r)=0$.
In Fig.\ref{regionplot2}, we plotted the location of the photon orbits and event horizon of an extremal
asymptotically AdS Kerr-Newman black hole with $\ell =m=1$. 
We are only interested in the region to the right of the event horizon (the exterior spacetime). 
This result is rather similar to the asymptotically flat case, even with a relatively small value of $\ell$ (i.e. relatively large $|\Lambda|$). That is to say, the cosmological constant does not affect the result found in \cite{Khoo:2016xqv} too much -- here we still observe that sufficiently small $a$ allows stable photon orbit on the extremal horizon, whereas large $a$ destabilizes the photon orbit on the extremal horizon.


\subsection{Case II: Kerr-AdS Black Hole}

\begin{figure}[h!]
\centering%
\includegraphics[width=.60\textwidth]{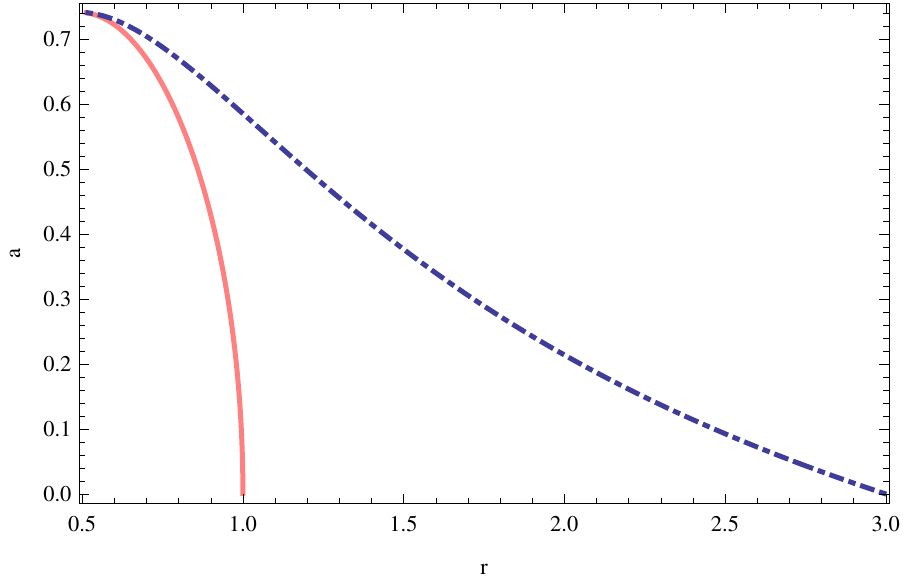} 
\caption{{} This plot shows the (outer) event horizon of a Kerr-AdS black hole (pink solid curve), and the prograde photon orbit (blue dashed curve). In the limit $a \to 0$, the dashed curve recovers the photon orbit for Schwarzschild-AdS black hole, which is located at $r=3m=3$. The photon orbit coincides with the event horizon for the extremal case.}
\label{AdSK}
\end{figure}

\begin{figure}[h!]
\centering%
\includegraphics[width=.60\textwidth]{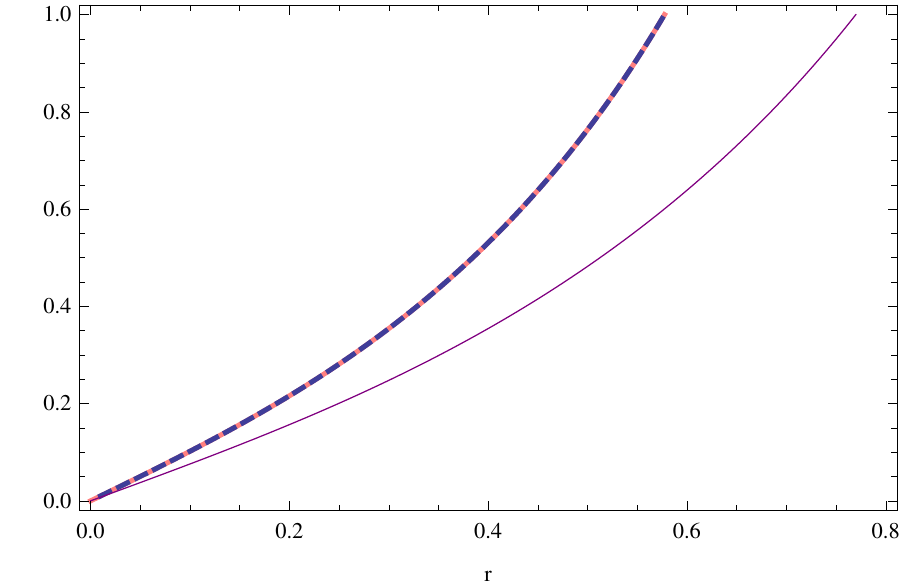} 
\caption{{} This plot shows the event horizon of an extremal Kerr-AdS black hole (solid red curve), and the photon orbit (dashed blue curve) which coincides with the horizon. The violet curve is the boundary of the stability region -- any photon orbit on the left of the curve is unstable.}
\label{AdSK2}
\end{figure}

The $e=0$ case corresponds to the Kerr-AdS black hole, which is worth a separate mention. In the extremal case, it turns out that there is indeed a photon orbit on the event horizon. Here we illustrate in Fig.(\ref{AdSK}) an example in which we have set $m=\ell=1$. This statement of course holds for all extremal Kerr-AdS black holes as we vary the mass, as shown in Fig.(\ref{AdSK2}). 

A simple and straightforward stability analysis shows that such a photon orbit on the extremal horizon is unstable, just like its asymptotically flat cousin. This is not surprising since we have shown previously that in the extremal Kerr-Newman AdS case, sufficiently large rotation (relative to the charge) destabilizes the photon orbit.




\section{Discussion: Extremal Black Holes Instabilities Galore}

We have set out to investigate the following question: Is the observation in \cite{Khoo:2016xqv}, namely that angular momentum tends to destabilize photon orbit on the event horizon (in the Doran sense) of an extremal black hole, whereas electrical charge tends to stabilize it, remains true in the presence of a negative cosmological constant? We found that the answer is affirmative, at least in the $(3+1)$-dimensional case. Of course, if the cosmological constant is small (i.e., $\ell$ is large), we would expect that we should recover the behavior observed in the asymptotically flat case. However, even with a relatively large cosmological constant, we still observe qualitatively the same behavior. Nevertheless, there are quantitative differences between the asymptotically AdS and the asymptotically flat cases. This is to be contrasted with the case of static black holes -- an asymptotically AdS Schwarzschild black hole (with spherical topology) possesses an unstable photon orbit at $r=3M$, which is identical to its asymptotically flat cousin.  

On an asymptotically $\text{AdS}_{3}$ background, an extremal charged non-rotating BTZ black hole has a stable photon orbit on its horizon, much like its $(3+1)$-dimensional Reissner-Nordstr\"om cousin.  For an extremal charged and rotating BTZ black hole with angular speed $|\omega|<\ell$, there is also a stable photon orbit on the horizon except for the special case $b=\omega$, in which no photon orbit exists. In the purely rotating case, BTZ black hole allows any $r=\text{const.}$ to be a photon orbit if $b=\ell$. This constant potential behavior appears to be a peculiar property of $(2+1)$-dimensions. These orbits are ``borderline'' unstable in the sense that any slight perturbation would still move the particle off its initial orbit. It does not admit any stable photon orbit on its horizon, just like the (3+1)-dimensional case.

For a charged and rotating BTZ black hole, we also observe that the geometry imposes an upper bound on the angular speed $|\omega|<\ell$, which is similar to the bound $|a| < \ell$ in the case of a Kerr-Newman-AdS black hole. However, unlike the latter case, the BTZ case always possesses a stable photon orbit on its horizon for all admissible value of the angular speed. That is to say, angular momentum never becomes large enough to destabilize the photon orbit on an extremal horizon in $(2+1)$-dimensions. 

Finally, we comment on the relationship between the presence of photon orbit on the extremal horizon and stability of the spacetime. There are two effects at play here: the first is simply due to the fact that if a photon orbit is stable, then a huge number of massless particles can pile up on the orbit, which causes a backreaction to the initial geometry -- the spacetime cannot be stable. This is a kind of dynamical instability. The end state of such an instability remains to be fully understood.

There also exists a less well known instability, which is of a thermodynamic type. More specifically, the presence of photon orbits, regardless of their stability, signals the possibility of a York-Hawking-Page type phase transition \cite{Y, HP}. This is due to the fact that the Dirichlet boundary-value problem in Euclidean quantum gravity sometimes has multiple solutions \cite{0301026}, which jump in numbers when the boundary passes through a photon orbit. Incidentally, it was observed in \cite{1610.01744} that there is a thermodynamical phase transition between an extremal charged (but non-rotating) BTZ black hole and a non-extremal one. In the same work, it was pointed out that there are many differences between the extremal case and the non-extremal case. In particular, an analysis of the quasi-normal modes showed that an extremal charged BTZ black hole is dynamically unstable under perturbation (see also the literature on Aretakis instability \cite{9808097, 1110.2006,1206.6598,1208.1437,1307.6800,1606.08505,1612.01562}). Indeed, it has been observed that in many examples, dynamical instability revealed from quasi-normal modes is a good indicator of thermodynamical instability, which involves a phase transition \cite{0703102, 0712.0645, 1405.2644}. 
These are all consistent with our result that an extremal charged BTZ black hole admits stable photon orbit on its event horizon, whereas its non-extremal counterparts do not. 

Finally, we emphasize that extremal black holes could exhibit properties that are absent in the non-extremal black holes, no matter how close to extremality they may be. After all, the process of taking limits in spacetimes is highly non-trivial \cite{carroll,geroch,ingemar}.  We should therefore not be alarmed that an extremal static black hole (satisfying certain conditions, see \cite{Khoo:2016xqv}) allows a stable orbit on its horizon. In general, when rotation is involved, the question of whether the photon orbit is actually on the horizon itself becomes subtle \cite{1107.5081, 9910099}. The stability of this orbit depends on the charge-to-angular momentum ratio, as well as the spacetime dimensions. The relationship between photon orbits, quasi-normal modes, dynamical instability and thermodynamical instability, is an interesting problem that has received quite a lot of attention lately \cite{1609.00083,1705.05461,1705.05928}, and should be further investigated.

\section*{Acknowledgement}

The authors acknowledge the support by the National Natural Science Foundation of China (NNSFC). Yen Chin Ong thanks Brett McInnes for useful discussions. He also thanks Fech Scen Khoo for comments and suggestions. The authors thank an anonymous reviewer for pointing out some technical errors in the the first version of the draft.


\end{document}